\documentstyle[11pt]{article}
\begin{document}
%\preprint{SUNY-FRE-98-02}
%\renewcommand{\baselinestretch}{1.2}
\title{Monopole Loop Distribution and Confinement 
in SU(2) Lattice Gauge Theory}
\author{Michael Grady\\
Department of Physics, SUNY College at Fredonia, 
Fredonia NY 14063 USA}
\date{\today}
\maketitle
\thispagestyle{empty}
\begin{abstract}
The abelian-projected 
monopole loop distribution is extracted from
maximal 
abelian gauge simulations.  
The number of loops of a given length 
falls as a power nearly independent
of lattice size.  This power increases with $\beta=4/g^2$, reaching
five around $\beta=2.85$, beyond which, it is shown, loops
any finite fraction of the lattice size vanish in the 
infinite
lattice limit. 
%\\ PACS:11.15.Ha, 11.30.Qc.
\end{abstract}

A strong correlation has been established between confinement and 
abelian monopoles extracted in the maximal abelian gauge.  
The entire SU(2) string tension appears to be due to the 
monopole portion of the projected U(1) field\cite{klsw,ekms}. 
Confinement appears to 
require the presence of large loops of monopole current of order the
lattice size, possibly in a percolating cluster.  As 
$\beta=4/g^2$ is raised, 
the finite lattice theory
undergoes a deconfining phase transition 
which is coincident with the 
disappearance of large monopole loops\cite{bmmp}.
This transition is interpreted as a 
finite-temperature phase transition,
one which exists if one of the four lattice dimensions is kept
finite as the others become infinite, but which disappears in the 
4-d symmetric infinite lattice limit.

In the U(1) lattice gauge theory itself, monopoles have also been
identified as the cause of the phase
transition\cite{stack}.  However, in this theory, 
monopoles are interpreted
as strong-coupling lattice artifacts which do not survive the 
continuum limit. The continuum limit is non-confining. 

Assuming the presence of large monopole loops is a necessary condition
for confinement,
in order for the SU(2) theory to confine in the continuum limit
it is necessary for some abelian monopoles to survive
this limit, i.e. to exist as physical objects. Evidence has 
been presented of scaling of the monopole density which
would make this so\cite{mdnz}. However, 
it is not sufficient just to have 
some monopoles survive this limit; one needs 
large loops of a finite physical size
to survive.  Small loops of finite size on the lattice, 
which are by far the most 
abundant, will shrink to zero physical size as the lattice spacing
goes to zero, becoming irrelevant. 
It is believed that to cause confinement, monopole loops
must be at least as large as the relevant Wilson loops of nuclear
size, and may need to span the entire space.

Consider
a large but finite universe, represented as an
$N^4$ lattice with lattice spacing $a$. 
The strong interactions 
should not care whether the universe is finite or infinite, so 
long as it is much larger than a hadron. One can then take the 
continuum limit as $a \rightarrow 0$, $N \rightarrow \infty$,
simultaneously, holding $Na$ constant at the universe size. It 
is then clear that the size of any object that is to remain of finite
physical size in the continuum limit must also become infinitely
large in lattice units, with linear dimension proportional to $N$,
i.e., some finite fraction of the full lattice size. It would therefore
appear that at the very least, monopole loops 
some finite fraction
of the lattice size must survive the continuum limit if it is to be 
confining, and probably loops at least as large as the lattice itself.   
Recently, it was shown that if the plaquette is restricted
to be greater than 0.5, the loop distribution function falls
so fast that no monopole loops any finite fraction of the lattice
size survive in the large lattice 
limit for any value of $\beta$\cite{loop}.
Here it will be shown that for the standard Wilson action
the same is true if $\beta > 2.85$.

By studying the monopole loop size distribution function,
one can tell how quickly the probability of finding loops
of increasing size decreases with loop size. 
This function turns out to have
very little finite lattice size dependence
for loops of length 
less than twice the lattice size.
Thus one can fairly confidently extrapolate this function, which
appears to be a simple power law, to the infinite lattice, from
which the probability of finding loops of finite physical size 
can be determined. Due to the dependence of the power 
on $\beta$, it is shown below that the probability
of finding loops whose length is a  finite 
fraction of the lattice size
vanishes in the large lattice limit for all $\beta > 2.85$.

Gauge configurations from the 
simulations were transformed
to maximal abelian gauge using the adjoint 
field technique\cite{haymaker}. 
Abelian monopole currents were then extracted 
using the DeGrand-Toussaint
procedure\cite{dg}. 
Sample sizes ranged from $500$ configurations 
for the $20^4$ lattices (1500 at $\beta = 3.1$), to $5000$ for the 
$8^4$ lattices,	always after 1000 equilibration sweeps.

Define the 
loop distribution function,
$p(l)$, as
the probability,
normalized per lattice site, of finding a monopole loop 
of length $l$ on 
a lattice of any size. 
The probability of finding a loop 
of length $N$ {\em or larger} on an 
$N^4$ lattice is given by $N^4 I(N)$ where $I$ 
is the integrated loop distribution
function
\begin{equation} I(M)=\int_{M}^{8N^{4}} p(l) dl ,
\end{equation}
where, since $N$ will be taken large, the 
discrete distribution has been
replaced by a continuous one.
The upper limit comes from the fact that each dual link
can contain at most two abelian monopoles.
To get confinement from the monopole mechanism, at least some 
finite fraction of lattices would
have to contain loops of length  order $N$ or larger. 
(Some would
argue that loops of size $N^2$ or larger might be 
necessary due to loop 
crumpling). Conversely, if 
\begin{equation}
\lim_{N\rightarrow  \infty} N^4 I(N) = 0 \end{equation}
then there will be no loops of length $N$ 
or larger on the $N^4$ lattice
in the large lattice limit, 
and 
confinement from the monopole mechanism will not be possible.

In Fig.~1, $\log_{10} p(l)$ is 
plotted vs. $\log_{10}(l)$ for various
$\beta$ and lattice sizes. 
The data are consistent with a power law, $p(l) \propto l^{-q}$, 
for loops up to around length $l=1.5N$ 
(the size 4 loops, which fall well below the 
trend are excluded from the fits).
This is consistent with the results of \cite{teper}
where the loop distribution was studied over a narrower $\beta$ range.
The larger
the lattice, the further the power law 
is valid before some deviation occurs
at large $l$. 
Also note
that the $12^4$ and $20^4$ data are virtually 
identical for loops up to size 30 or so, and even
the $8^4$ are hardly different.
Linear fits were made for loop sizes in the range 6 to 1.1N
(only
the $12^4$ fits are shown for clarity).
For larger loop sizes, occasionally zero instances
of a particular size was observed. Zeros cannot be
plotted on the logarithmic scale, 
but if ignored the data will be
biased upward. A moving average was used
in this circumstance to properly account for the 
zero observations.   

The deviations from linearity for large 
loops can be easily understood as a finite size effect
coupled with the periodic boundary condition. For 
loops longer than about $1.5N$,
there is a significant probability of reconnection through 
the boundary.
This makes a would-be large loop terminate earlier 
than it would on 
an infinite lattice. Thus, on a finite lattice there 
must be a deficit of
very large loops, and an {\em excess} of 
mid-size loops due to 
reconnection.   
At $\beta=2.4$ and $2.5$ this
is especially apparent.
The linear trend continues further for the $20^4$ lattice
than for the $12^4$, which itself
continues further than the $8^4$.
At some point starting around $1.5N$, but not readily
apparent until about $4N$,
there is a bulge of excess probability, 
followed eventually by a steep
drop that falls below the linear trendline.
Often the data cuts off before it falls below,
but the fact that no loops larger than those plotted occurred can be
used to infer that the probability distribution must eventually
fall below the trendline.
For instance if one assumes that very large loops
follow the trendline
for the $12^4$ data at $\beta = 2.4$, 
then one can calculate that 3.5 instances
of loops longer than  1250 lattice units should have been seen in
the sample. The fact than none occurred implies 
that the data most likely does fall below the trend line for very
large loops, and certainly could not remain significantly above it.	 

Because the power 
law trend continues further
the larger the lattice 
and the deviations always occur for $l > N$, 
it seems quite reasonable to assume that on the 
infinite lattice one would
have a pure power law. 
It is difficult to imagine what length other than the lattice size
could set the scale for 
a change in behavior at extremely large loop sizes beyond 
those measured here. 
In addition, since the small loop data are
nearly independent of lattice size, it would seem 
reasonable that 
the power, which is completely determined by the small loops, 
must also be essentially the same
on the infinite lattice as observed here for 
the $12^4$ or $20^4$ lattices (some slight variations 
are visible on the 
$8^4$).  
Assuming this, one can easily predict the point at 
which condition (2) 
becomes satisfied, namely $q>5$. For $q>5$ the 
probability of having
a monopole loop with length equal 
to {\em any finite fraction} of the lattice
size $N$, or larger, vanishes in the large lattice limit, 
whereas for $q<5$ the same 
becomes overwhelmingly likely. 
This is because for $q>5$ not only is condition (2) satisfied, but
also \begin{equation}
\lim_{N \rightarrow \infty} N^{4} I(N/M) = 0 \end{equation}
for any finite $M$.

The power $q$ is plotted 
as a function of $\beta$ in Fig.~2. A definite
rising trend is observed, in sharp contradiction to the 
conclusion of ref. \cite{teper}, where a constant value of $q$
was inferred from runs over a rather narrow $\beta$ range (2.3 to 2.5).
It is seen that $q$ apparently passes 5 around $\beta=2.85$. 
For any $\beta$ beyond this, including the 
continuum limit, $\beta \rightarrow \infty$, there will be no loops
any finite fraction of the lattice size in the infinite lattice limit.
A small residual finite size 
dependence in $q$ is possible,
which could shift the infinite lattice 
critical $\beta$ to 2.9 or possibly 3.0, but it is hard to picture 
how this could prevent $q$ from ever reaching 5, which would be 
necessary to have large loops survive the continuum limit.

Error bars in Fig.~2 were obtained from the least squares fits.
If data for different loop sizes were highly correlated, then
these errors could be underestimated. As a test, 
correlation coefficients
were computed between the numbers of each size loop occurring on a 
lattice,
for samples of 1000 $12^4$ lattices at both $\beta=2.6$ and 2.9. The 
degree
of correlation was in all cases very small, with coefficients, $r$,
averaging 0.02 (with a maximum of 0.04)
at $\beta=2.6$ and 0.003 at $\beta=2.9$. These fall 
within
the expected noise level of 0.03 for these sample sizes, and are 
certainly small
enough to justify ignoring correlations in the fits. The 
autocorrelation
in Monte Carlo time was also computed and found to fall below noise
after two Monte Carlo sweeps. One might also ask with what confidence
one can {\em reject} the hypothesis $q \le 5$ at $\beta=3.0$ or $3.1$. 
At  $\beta=3.1$ on the $20^4$ lattice, q was found to be 
$7.65 \pm 0.32$.  
If one assumes $q=5$ instead, then the number of loops of size
12 links and larger that should have occurred in the sample
(based on the number of
six-link loops) is 285, whereas the actual number was 35. 
The possibility
of this occurring is, 
by raw Poisson statistics, less than $10^{-76}$. At
$\beta=3.0$ the same analysis rejects 
the hypothesis at the level $10^{-59}$.
This shows that on these lattices it is 
virtually certain that $q$ exceeds 5,
even if some latitude is allowed for the 
modest Monte Carlo time correlations.

It is interesting to note that the 
above conclusions are insensitive to the 
precise behavior of the distribution
for $l > N$. If $p(l)$ 
follows a different power law
with exponent $q'$ for $l>N$, $q'$ 
could be as small as {\em unity} and
still condition (2) will hold. For instance, if $q=5+\epsilon$ and
$q' = 1$ then 

\begin{equation}
p(l) = \left\{ \begin{array}{lr}
l^{-5-\epsilon} &  \mbox{, } l \leq N \\
N^{-4-\epsilon} l^{-1}    &  \mbox{, } l > N  \end{array}
\right. ,\end{equation}
so \begin{equation} 
I(N) \propto N^{-4-\epsilon} \int_{N}^{8N^{4}} l^{-1} dl ,
\end{equation}
and $N^4 I(N) \propto N^{-\epsilon} \ln (8N^{3})$ which
still vanishes as $ N \rightarrow \infty$. Thus the presence
of large enough loops to cause confinement is, paradoxically, 
controlled almost entirely by the distribution function for 
loops of length smaller than $N$, because this determines the
base probability for the case $l=N$.  This observation makes the 
above result far more robust, since the distribution for $l<N$
is seen to always follow a pure power law which is almost independent
of lattice size. 
It was the distribution for $l>N$ which was somewhat less certain, but
all that one needs to know about this part of the distribution 
is that it falls at least as fast as $l^{-1}$, 
not a very stringent requirement. 
As before, the loop size cutoff for the condition 
can be taken to be $N/M$ instead
of $N$, where M is some finite number. This further
reduces the likelihood of 
the finite lattice size affecting
the important small-loop part of the distribution.

On our lattices, as in many previous studies, confinement
appears to occur at couplings for which
loops that wrap completely around
the periodic
lattice are common, and the theory is deconfined when
such loops are absent.  
This is illustrated in Fig.~3, where the probability that the
largest loop has a non-zero winding number is plotted vs. $\beta$,
along with the Polyakov loop. The rather sudden rise from zero in
winding probability is coincident with the point of largest slope
in the Polyakov loop, which signals deconfinement. For the $20^4$
lattice, the average Polyakov loop itself 
is not a very sensitive test of confinement,
but moments of the Polyakov loop become decidedly non-Gaussian
at this point. For instance both the first (absolute value)
and fourth moment
are consistent with Gaussian for $\beta \leq 2.7$, but
not for $\beta > 2.7$ where the distribution is not peaked at zero and
the theory is deconfined. 
If loops in addition to the largest loop were 
also tested, the net
winding probability would likely 
become very close to unity when
deeply into the confining region.
Loops which wind the lattice occur in pairs, with a Dirac sheet
stretched between them. Once a winding pair exists, 
the loops can drift
apart without increasing their length, stretching the Dirac sheet
to cover an arbitrarily large section of lattice and incurring
no additional energy penalty in doing so. 
Such a Dirac sheet could easily
disrupt the values of Polyakov loops 
passing through it only once, 
resulting in 
a random, i.e. confined, Polyakov loop average.
Very large non-winding loops 
would also have large Dirac sheets having
the same effect. However, 
the winding configurations, having shorter loops,
have a lower energy, so 
would appear first at the periodic lattice 
deconfinement transition.

Since large loops must disappear when $q$ hits 5, if one accepts
the observation that large loops are associated with confinement,
one would have to conclude that
the infinite 4-d symmetric lattice 
must undergo a deconfining phase transition at a
finite value of $\beta$ around 2.9, 
rather than at $\beta=	\infty$
as is usually assumed. 
Of course, one could 
instead give up the link between abelian monopoles and confinement,
despite the strong evidence in favor of a connection.
Since large abelian monopole loops 
have been shown to be responsible for most if
not all of the string tension, if the theory still confines when
they are absent it will have a
much smaller string tension. 
In this case the theory would confine in the continuum, but 
could be quite different in detail from the usual lattice theory
in the crossover region of the Wilson action, 
since the latter would be highly contaminated with monopoles,
seen here as essentially lattice artifacts.

Let us return to the possibility that the continuum 
non-abelian pure-gauge theory really doesn't confine,
an idea which has been suggested before\cite{zp,ps2}. 
The renormalized coupling could have an infrared-stable fixed
point, and stop increasing at some distance scale. 
The heavy-quark
potential would be a logarithmically-modified 
Coulomb potential,
and the gluons would be massless. 
How is one then to explain confinement
in the real world? It 
may be that the real-world confinement that 
we see is a manifestation of chiral symmetry 
breaking\cite{zp,nc}. Even in
a nonconfining theory, chiral symmetry can break 
spontaneously if the 
coupling is strong enough. Instantons will
also likely play a role here. 
Once a chiral condensate is established,
it can produce a confining force by 
expelling strong color fields;
i.e. if the condensate abhors external 
color fields, hadrons
will expel some condensate from their immediate
vicinity, creating a region of higher 
vacuum energy proportional
to their volume, a sort of chiral-expelled 
bag. Stretching of this
bag would result in a linear potential. 
The energy density of the bag
which would follow the movement of quarks 
may also be responsible for the
quarks' dynamical mass.
This picture is supported by
the observation that $<\! \bar{\psi}\psi \! >$ is 
lowered in the 
neighborhood of a color source\cite{markum}, 
indicating some 
expulsion of 
condensate. 
Similar ideas are contained in chiral
quark models\cite{cqm} where a 
polarized Dirac sea is responsible 
for the binding of the quarks in a baryon, 
and also in the 
instanton liquid model\cite{ilm}. 
Both of these models
are able to compute with fair accuracy a large number of 
low-energy properties of hadrons, and neither has an
absolutely confining 
potential.  Further details on this scenario 
and justifications are given in \cite{loop}
and references therein.
\section*{Acknowledgement}
It is a pleasure to thank 
Richard Haymaker for advice on implementing
the adjoint field technique.

\newpage
\begin{center}
{\Large Figure Captions}
\end{center}
\noindent
FIG. 1. Log-log plots of loop probability vs. loop
length: (a) $\beta=2.4 $ (upper graph, right scale), 
$\beta=2.5$ (left scale, 
shifted for clarity);
(b) from upper to lower, data series are $\beta=$ 2.6, 2.7, 
2.8, 2.9, 3.0, 3.1;  linear fits are to upper region of the $12^4$ 
data, as explained in text. \\\\

\noindent
FIG. 2. Power, $q$, characterizing the decrease of loop 
probability with
loop length, vs. $\beta$. 
Lines are drawn to guide the eye. \\\\

\noindent
FIG.3. Polyakov loop (filled symbols) and
winding probability of largest loop (open symbols) vs. $\beta$.
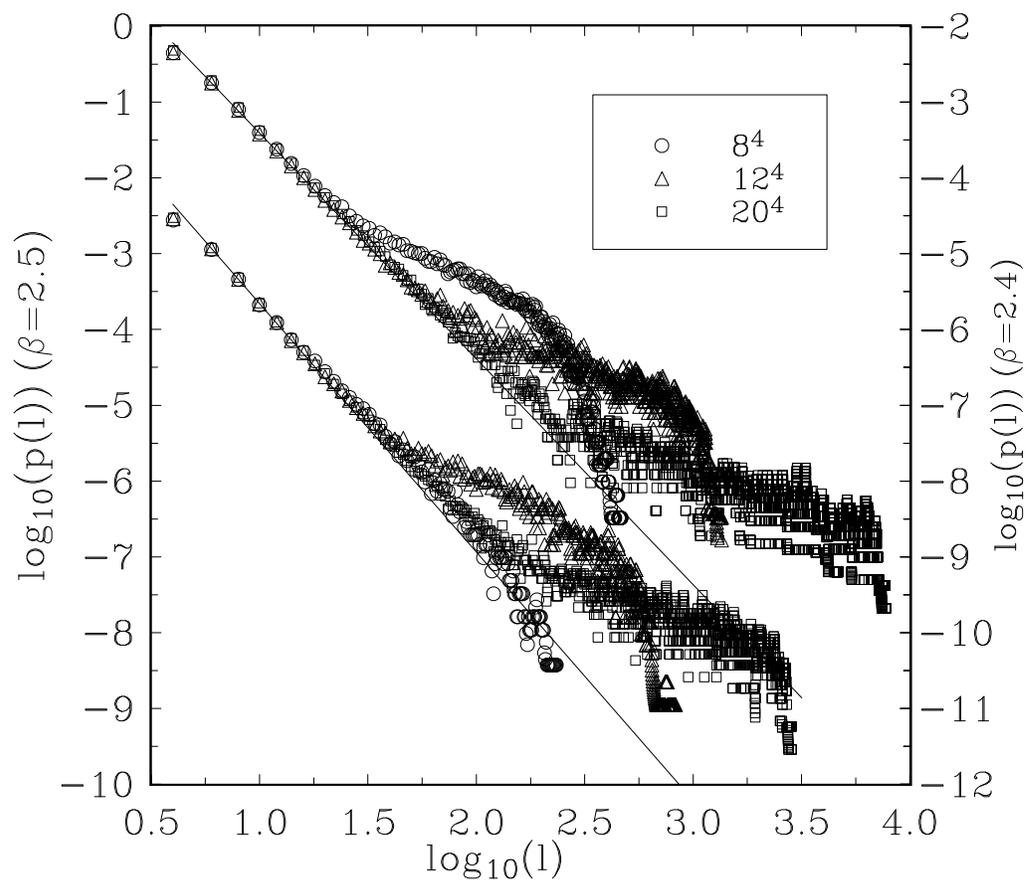
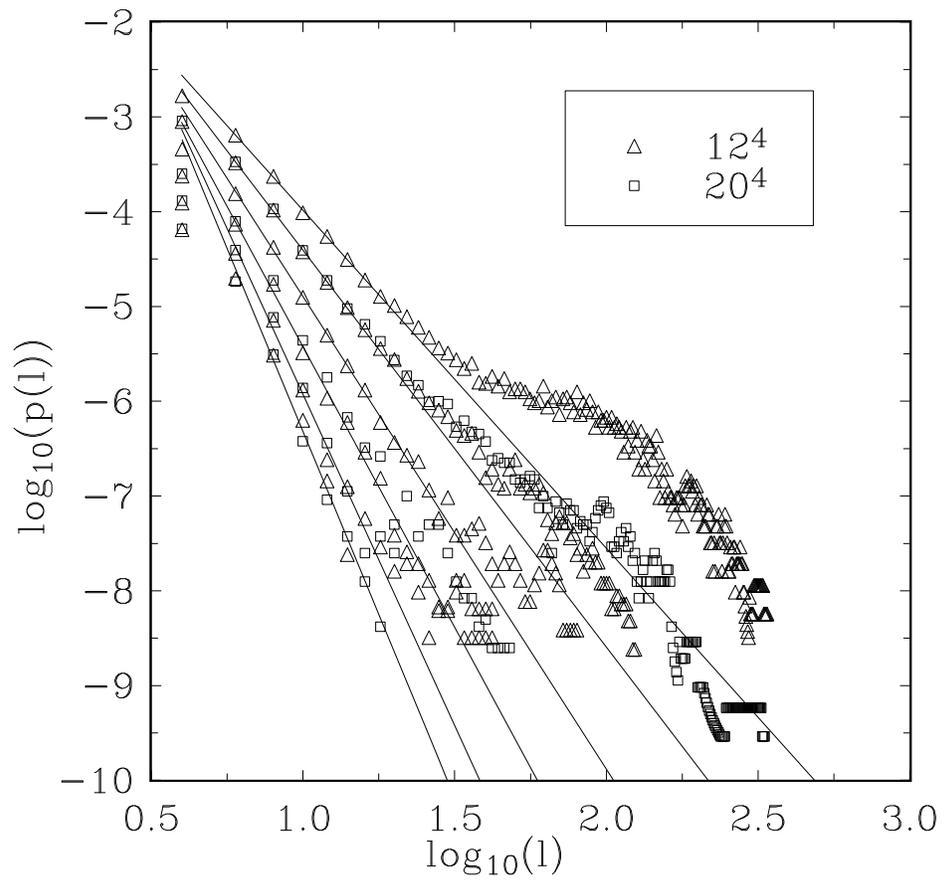
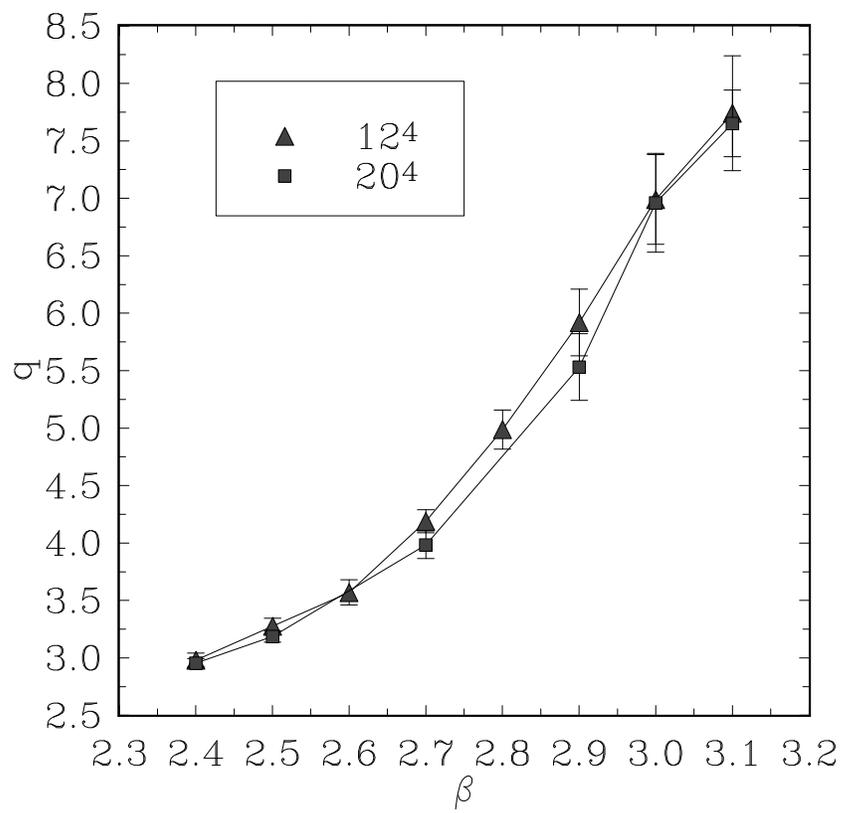
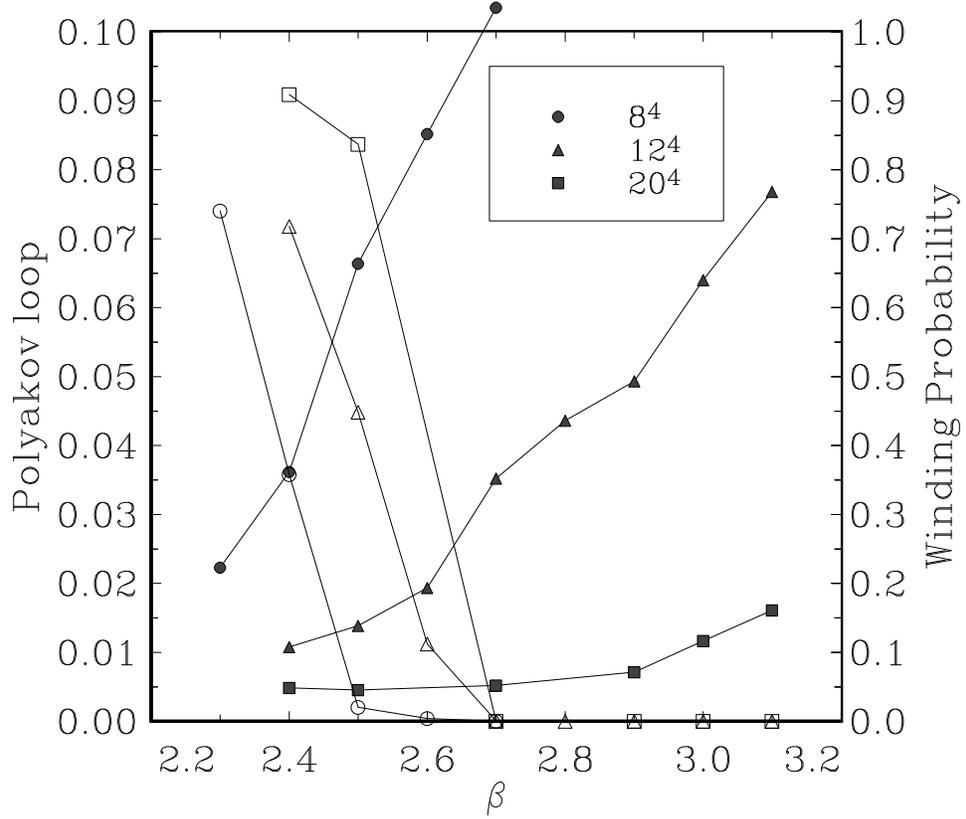
\end{document}